\documentclass[11pt]{article}
\usepackage{epsfig}
\usepackage{epstopdf}
\usepackage{amsmath}\usepackage{amssymb}
\usepackage{mathrsfs}
\usepackage{graphicx}

\newcommand {\be}{\begin{equation}}
\newcommand {\ee}{\end{equation}}
\newcommand {\ba}{\begin{eqnarray}}
\newcommand {\ea}{\end{eqnarray}}

\begin{document}

\def \a'{\alpha'}
\baselineskip 0.65 cm
\begin{flushright}
\ \today
\end{flushright}

\begin{center}{\large
{\bf Top pair Asymmetries at Hadron colliders with general $Z'$
couplings}} {\vskip 0.5 cm} {\bf Seyed Yaser Ayazi}{\vskip 0.5 cm
} School of Particles and Accelerators, Institute for Research in
Fundamental Sciences (IPM), P.O. Box 19395-5531, Tehran, Iran
\end{center}

\begin{abstract}
Recently it has  been shown that measurement of charge asymmetry
of top pair production at LHC excludes any flavor violating $Z'$
vector gauge boson that could explain Tevatron forward-backward
asymmetry (FBA). We consider the general form of a $Z'$ gauge
boson including left-handed, right-handed vector and tensor
couplings to examine FBA and charge asymmetry. To evaluate top
pair asymmetries at Tevatron and LHC, we consider $B^0_q$ mixing
constraints on flavor changing $Z'$ couplings and show that this
model still explain forward-backward asymmetry at Tevatron and
charge asymmetry can not exclude it in part of parameters space.

\end{abstract}

\section{Introduction}
The top quark is the only Standard Model (SM) particle which its
mass is at the order of the electroweak symmetry breaking and its
lifetime is very short. This feature causes that it decays before
it can form any hadronic bound state. Thanks to these particular
features, careful measurement of top quark properties may be
sensitive to new physics (NP).

Experimental results for the cross section of top pair at
Tevatron and LHC are well consistent with the SM prediction. While
waiting for discovery of NP at the LHC, CDF and D0
collaborations report deviation from SM prediction in the FBA in
top pair production \cite{CDF,Afb}. Actually, this observation at
Fermilab Tevatron may already be a hint of NP. In the SM, top
pair production can be produced via the $q\bar{q}$ annihilation
and $gg$-fusion. The interference between radiative corrections
involving gluon emission and box diagrams lead to FBA in the top
pair production \cite{Kuhn:1998kw}.

 Many extensions of SM have been proposed to
explain the measured FBA. Some of these models propose unknown
heavy particles which can be exchanged in top pair production
process \cite{AguilarSaavedra:2012ma}. Possible new particles
which can contribute to $t\bar{t}$ production are flavor violating
$Z'$ \cite{Z'}, $W'$ vector boson \cite{Ayazi:2012bb}, spin-2
boson \cite{Grinstein:2012pn}, axigluons \cite{axigluon}, twin
Higgs model \cite{twin Higgs model}, colored Kaluza Klein
excitations of gluon in warped Ads space \cite{KK gluon},
color-triplet scalar \cite{color-triplet scalar} and color-sextet
scalar \cite{color-tsextet scalar}.

The LHC allows us to investigate the properties of the top quark
in details. A very large number of top quarks are produced at the
LHC eventually more than $10^7$ $t\overline{t}$ pairs per year
\cite{Halkiadakis:2010mj}. This will make feasible the precise
investigations of the top interactions. Since the initial state of
proton-proton collisions at the LHC is symmetric, charge
asymmetry (CA) manifestly is different from FBA \cite{charge SM}.
CA at the LHC is defined as the difference between events with
positive and negative absolute values of rapidities of top and
antitop quarks. The CMS collaboration has recently presented CA
measurement in $t\bar{t}$ production at the LHC for the center of
mass energy $7~\rm TeV$ and $1.09~\rm fm^{-1}$ of data. This
measurement is well consistent with the SM prediction.
Nevertheless, the measurement of CA at LHC can provide an
independent criterion of NP models which explain FBA at Tevatron.

One of the models which can explain FBA anomaly at Tevatron  is
flavor violating $Z'$ vector boson. Recently, it has been shown
\cite{Fajfer:2012si}, that the LHC CA measurement exclude flavor
violating $Z'$ vector boson which explain the Tevatron FBA.
However, In this paper, we consider general form of $Z'$ flavor
boson which includes right-handed, left-handed vector and tensor
terms. We study the effect of this model on observables FBA, CA,
and total cross section of $t\bar{t}$ at the Tevatron and LHC. The
main point is that coupling of flavor violating $Z'$ exchange can
contribute to $B_q$ ($q=s,d$) mixing. For this reason, to study
the impact of $tuZ'$ vertex to top pair asymmetry, we will
consider $B_q$ constraints on these couplings.

The rest of this paper is organized as follows: In the next
section, we introduce $Z'$ boson with general coupling and its
effect on the cross section production of top-antitop. In
section~3 we summarize  observables which  we study at the LHC and
Tevatron and study the effects of flavor violating $Z'$ boson with
general coupling on our observables.  The conclusions are given
in section~4.

\section{Flavor Violating $Z'$ boson with general coupling}
In this section, we will focus on the model with $Z'$ gauge boson
that has general coupling and  briefly describe the model and
phenomenological constraints on its parameters space.

One of the extensions of SM has been proposed to resolve
discrepancy of FB asymmetry measured by CDF and $D0$
collaborations is the flavor violating $Z'$ vector gauge boson.
However, recently \cite{Fajfer:2012si} it has been shown that the
measurements of CA at LHC have excluded any $Z'$ vector gauge
boson. In this paper, we consider similar extension of SM with
$Z'$ gauge boson which has a general coupling including vector,
axial vector and tensor coupling. Since a tree level $dbZ'$ and
$sbZ'$ coupling contribute to $B^0_{d,s}$, we ignore these terms
in Lagrangian. The most general Lagrangian for flavor changing
$tqZ'$ ($q=u,c,t$) transition has been given by
\cite{Duraisamy:2010pt}:
\begin{eqnarray}
{\cal{L}}_{tuZ'} & =
\bar{u}[\gamma^{\mu}(a+b\gamma_5)+i\frac{\sigma_{\mu\nu}}{m_t}q^{\nu}(c+d\gamma_5)]t
Z'_{\mu},\label{exp}
\end{eqnarray}
where $ a, b , c$ and $d$ are real constants and $q=p_t-p_u$. This
Lagrangian can contribute to $t\bar{t}$ production at hadron
colliders via t-channel exchange of the $Z'$ boson.

This Lagrangian with large flavor changing up type quarks
contribute to FCNC processes and $B^0_q$ mixing. For instance,
$tuZ'$ coupling will generate a $bq'Z'$ ($q'=d,s$) coupling at
loop level which contribute to $B^0_q$ mixing. As a result,
$B^0_q$ mixing constrain the $tq'Z'$ coupling. Constraints on
$tq'Z'$ coupling have been estimated in \cite{Duraisamy:2011pt}.
To calculate FBA and CA, we consider these constraints on the
couplings. Note that right-handed $tuZ'$ coupling do not
contribute to $B^0_q$ mixing in the limit that up quark mass set
to zero. Also in the limit of low energy, the effect of tensor
coupling ($c$ and $d$) on $B^0_q$ mixing suppressed by $\sim
m_b/m_t$ and consequently, we can ignore the $B^0_q$ mixing
constraints on these couplings. In the following, we introduce
top pair production observables at hadron colliders and consider
the above coupling effects on them.

\section{Observables and Numerical results}
In this section, we study the total cross section of top pair
production at the Tevatron and LHC and consider top pair forward
-backward asymmetry  and charge asymmetry as observables and
study effect of $Z'$ gauge boson on them.

Top pair production cross section at Tevatron ($\sqrt{s}= 1.96
~\rm TeV$) has been measured by $\rm D0$ collaboration  with
$5.4~\rm fb^{-1}$ of integrated luminosity \cite{cross Tev}:
\begin{eqnarray}
\sigma_{\rm Tevatron}(pp\rightarrow t{\overline t} ) & =
7.56\pm0.83~[pb]~(\rm stat\oplus sys).\label{exp}
\end{eqnarray}
The cross section value for top pair production at LHC have been
measured by CMS experiment recently \cite{cross LHC}:
\begin{eqnarray}
\sigma_{\rm LHC}(pp\rightarrow t{\overline t} ) & =
165.8\pm13.3~[pb]~(\rm stat\oplus sys).\label{exp}
\end{eqnarray}
These measurements are in good  agreement with the SM prediction
\cite{SM Tev,SM LHC}. The tree-level total cross section for
$qq'\rightarrow t\bar{t}$ including both SM  and $Z'$
contribution has been calculated in \cite{Duraisamy:2011pt}. The
total cross section of top pair production at hadron colliders
can be obtained by convoluting the partonic cross section with
the parton distribution functions (PDF) for the initial hadrons.
To calculate $\sigma(pp\rightarrow t\overline{t})$, we have used
the CTEQ6L parton structure functions \cite{CTEQ} and set the
center-of-mass energy to $7~ \rm TeV$. The total cross section
for production of $t\bar{t}$ has the following form:
\begin{eqnarray}
\sigma(pp\rightarrow t{\overline {t}}) & = &\sum_{ab} \int
dx_1dx_2f_a(x_1,Q^2)f_b(x_2,Q^2) \widehat{\sigma}(ab\rightarrow
t{\overline {t}}), \
\end{eqnarray}
where $f_{a,b}(x_i,Q^2)$ are the parton structure functions of
proton. $x_1$ and $x_2$ are the parton momentum fractions and $Q$
is the factorization scale.

Here, we emphasis that for proton-antiproton collision FBA is
defined as relative difference between the number of produced top
quark with $\cos\theta >0$ and $\cos\theta<0$, which $\theta$ is
the production angle in the center of mass system:
\begin{eqnarray}
A_{FB} & =
&\frac{N_t(\cos\theta>0)-N_t(\cos\theta<0)}{N_t(\cos\theta>0)+N_t(\cos\theta<0)}\
\end{eqnarray}
As it is mentioned, SM model prediction for FBA is as small as a
few percent which arises from the interference between the Born
amplitude for $q\bar{q}\rightarrow Q\bar{Q}$ and box diagrams and
the interference term between initial state radiation and final
state radiation \cite{Kuhn:1998kw}. At the Tevatron, since the
initial state is asymmetric (proton-antiproton collisions), the
top quark forward- backward asymmetry can be measured. Recent
measurements by $\rm CDF$\cite{CDF} ($A_{FB}= 0.158\pm0.075$) and
$\rm D0$\cite{Afb} ($A_{FB}= 0.196\pm0.065$) collaborations
report deviation from the SM prediction which is about $2\sigma$
larger than the SM (value about $5\%$) predictions. At the LHC,
initial state is symmetric (proton-proton collisions), as a
result FBA vanishes. However, charge asymmetry in $t\bar{t}$
production at LHC can be measured which reflects the top quark
rapidity distribution. The top quark charge asymmetry in
$t\bar{t}$ is defined by \cite{Charge measurement}
\begin{eqnarray}
A_{C} & =
&\frac{N_t(\Delta(y^2)>0)-N_t(\Delta(y^2)<0)}{N_t(\Delta(y^2))>0+N_t(\Delta(y^2)<0)}\
\end{eqnarray}
where $\Delta(y^2)$ is defined as,
\begin{eqnarray} \Delta(y^2) &
= &(y_t-y_{\bar{t}}).(y_t+y_{\bar{t}})\
\end{eqnarray}
and $y_t$($y_{\bar{t}}$) are the rapidity of the top (anti)quark
in the laboratory frame. Rapidity difference is a boost invariant
observable and is equal to:
\begin{eqnarray}y_t-y_{\bar{t}} &= &
2\rm
Arc\tanh(\sqrt{1-\frac{4m^2_t}{\hat{s}}}\cos\theta)\end{eqnarray}
while summation of rapidities is not boost invariant and can be
written as:
\begin{eqnarray}
y_t+y_{\bar{t}}& = &\frac{1}{2}\ln(\frac{x_1}{x_2})\
\end{eqnarray}
In proton-proton collisions at the LHC, the rapidity distributions
of the top and antitop quarks are symmetrically distributed around
zero. But since the u, d valence quarks carry larger average
momentum fraction than the anti-quarks, $t\bar{t}$ boost  along
the direction of the incoming quark, and therefore this leads to
a larger average rapidity for top quarks than anti-top quarks.
The ATLAS and CMS measurements for the charge asymmetry are: $A_C
= -0.019 \pm0.036$ \cite{Ac Atlas}, $A_C =
-0.013\pm0.041$ ($A_C =
0.004\pm0.014$) \cite{Ac CMS} , and the SM prediction is $A_C =
0.0115$ \cite{Ac SM}. Notice that while measurements of the FBA
show a deviation from the SM expectations, measurement of charge
asymmetry at the LHC is in agreement with SM prediction. It means
that any new physics which explains the $t\bar{t}$
forward-backward asymmetry must satisfy $A_C$ measurements
consistent with the SM predictions. In the following, we study
general form of $Z'$ gauge boson and consistency with these
measurements.

\begin{figure}
\begin{center}
\centerline{\epsfig{figure=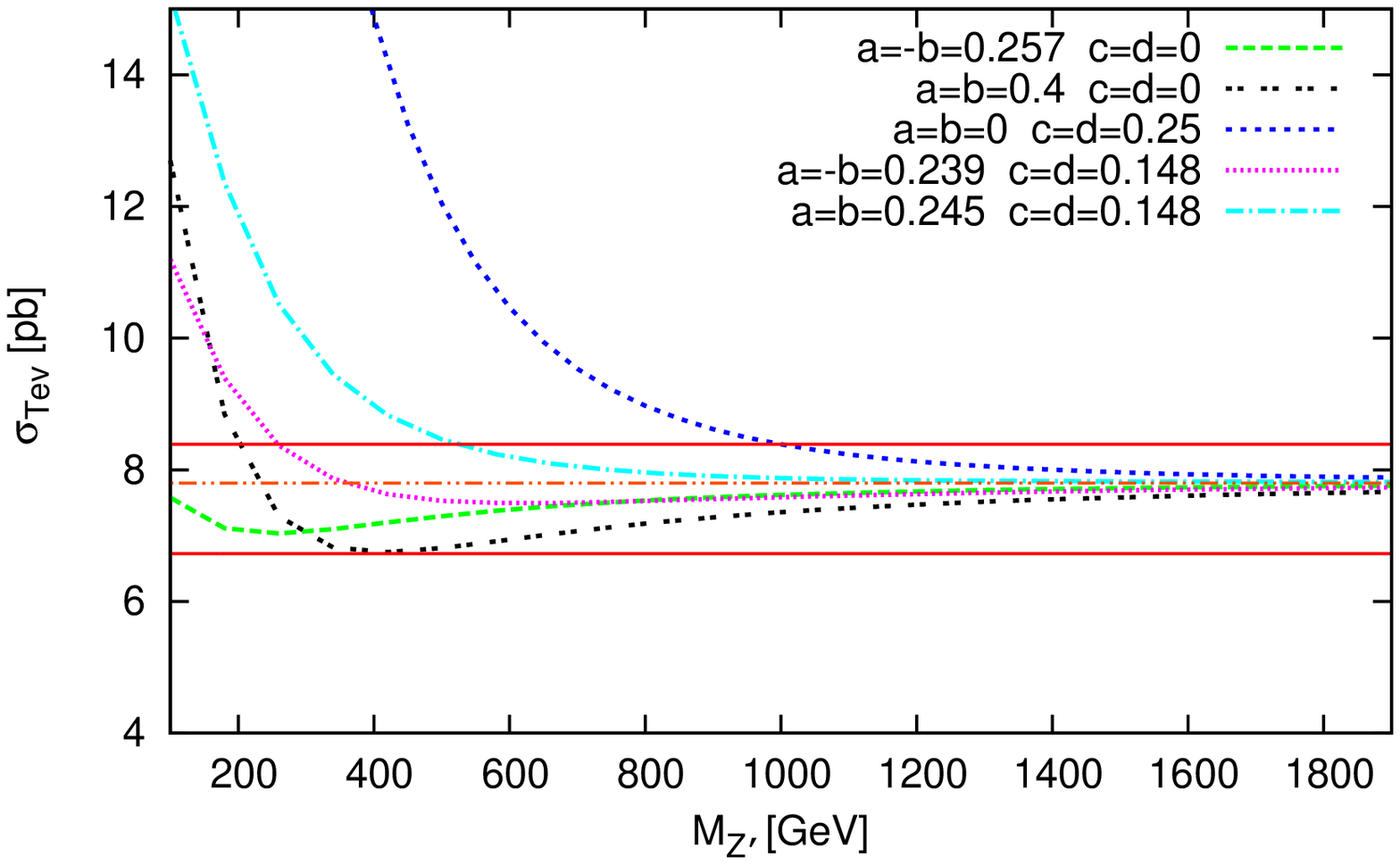,width=7.5cm}\hspace{5mm}\epsfig{figure=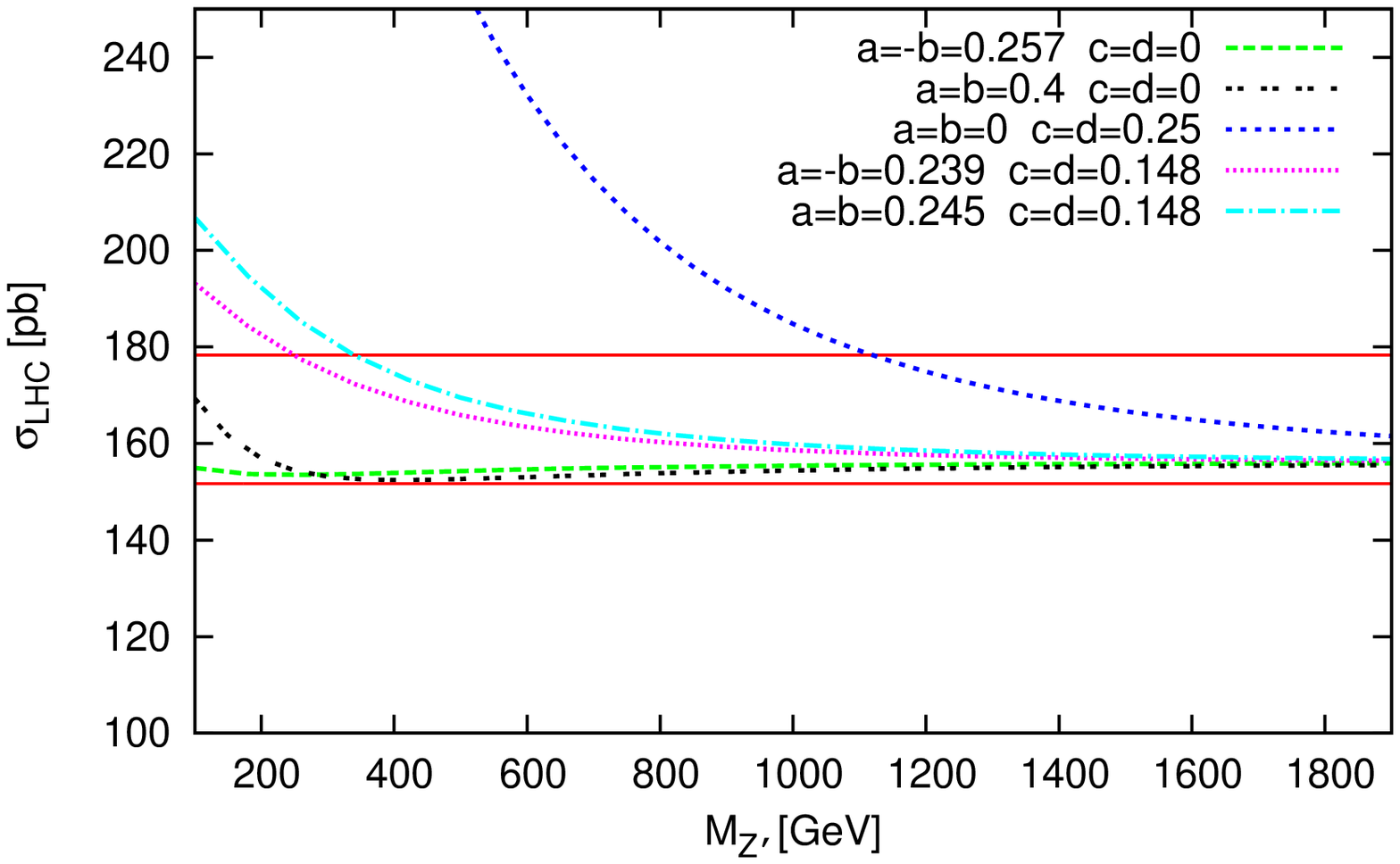,width=7.5cm}}
\centerline{\vspace{2.cm}\hspace{0.5cm}(a)\hspace{7cm}(b)}
\centerline{\vspace{-3.5cm}}
\end{center}
\caption{The top pair production cross section as a function of
the $Z'$ mass at Tevatron (a) and LHC (b). The horizontal red
lines show allowed range of experimental measurements for the top
pair total cross section. } \label{cross}
\end{figure}

In numerical calculation, we have set $m_t=172.5~\rm GeV$ and
fixed renormalization and factorization scale $\mu_R=\mu_F=m_t$.
For including higher order QCD effects, we have normalized all
observables to ratio of measured experimental cross section to
the leading order SM cross section. In Fig. \ref{cross}, we have
displayed  the total cross section of top-antitop production at
the Tevatron and LHC as a function of the $Z'$ mass. In this
figure, $\Gamma_{Z'}=2$~GeV and different values for couplings are
considered. As it is mentioned, due to contribution of $tqZ'$
coupling to $B^0_q$ mixing, the $\Delta M_q$ experimental results
can constrain these couplings. It is shown
\cite{Duraisamy:2011pt} that a global analysis on parameters of
mass mixing for $B^0_d$ and $B^0_s$ mixing constrains $a=-b=g_L$
coupling down to 0.4. For $tuZ'$ vertex with right-handed
coupling $a=b=g_R$, we can avoid the $B^0_q$ mixing due to
suppression  value $m^2_u/m^2_W$. Also for tensor couplings $c$
and $d$, the contribution of these operators to $B^0_q$ mixing at
low energy is negligible \cite{constraint}.

In Fig.~\ref{cross}, we consider all $tuZ'$ coupling constraints
which arise from $B^0_q$ mixing. The horizontal red lines show
the allowed range of experimental measurement for top pair total
cross section at the Tevatron and LHC. The curves with different
colors and lines show various values of coupling, according to
pseudo vector, vector, tensor and general form of $Z'$ couplings.

\begin{figure}
\begin{center}
\centerline{\epsfig{figure=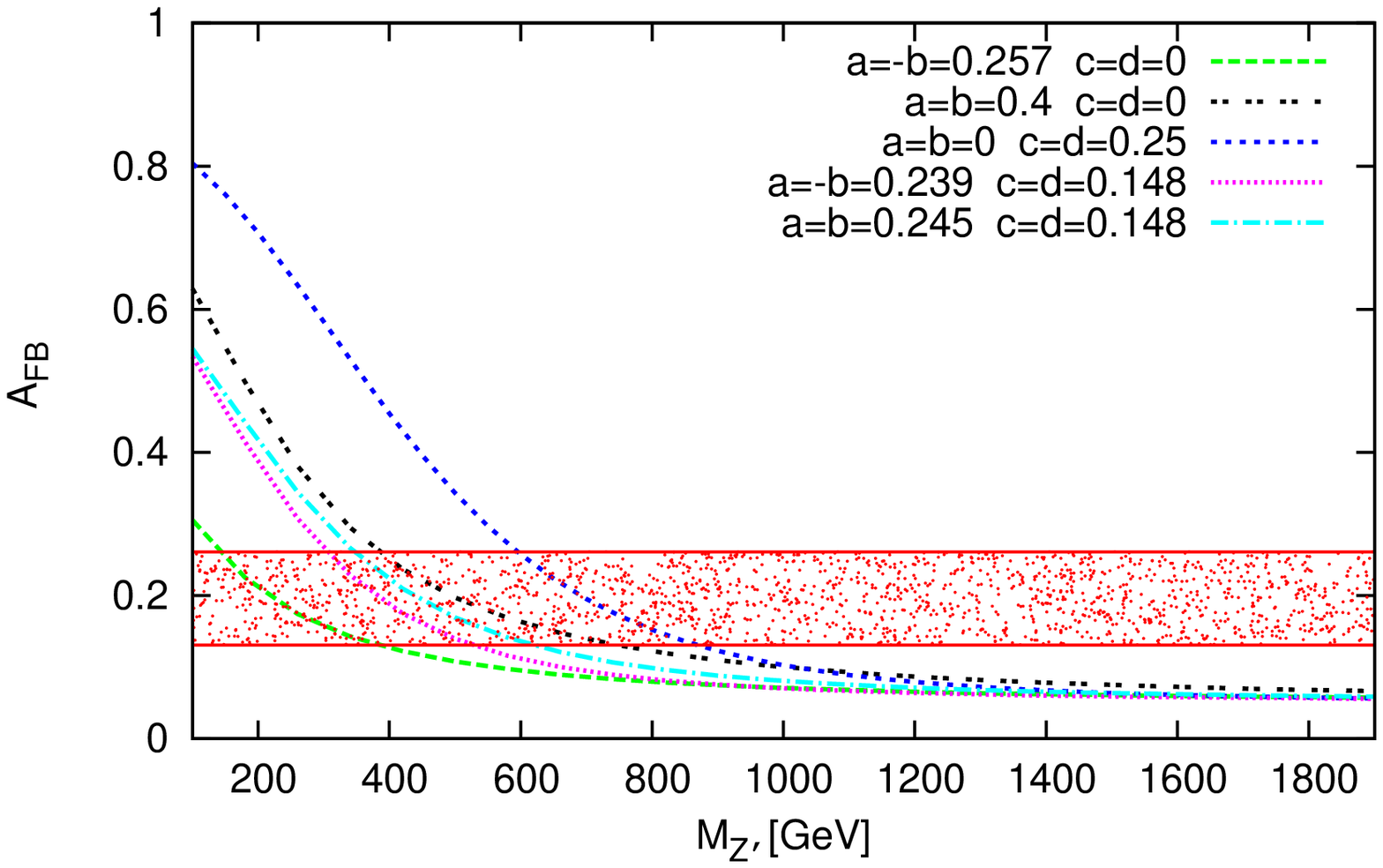,width=7.5cm}\hspace{5mm}\epsfig{figure=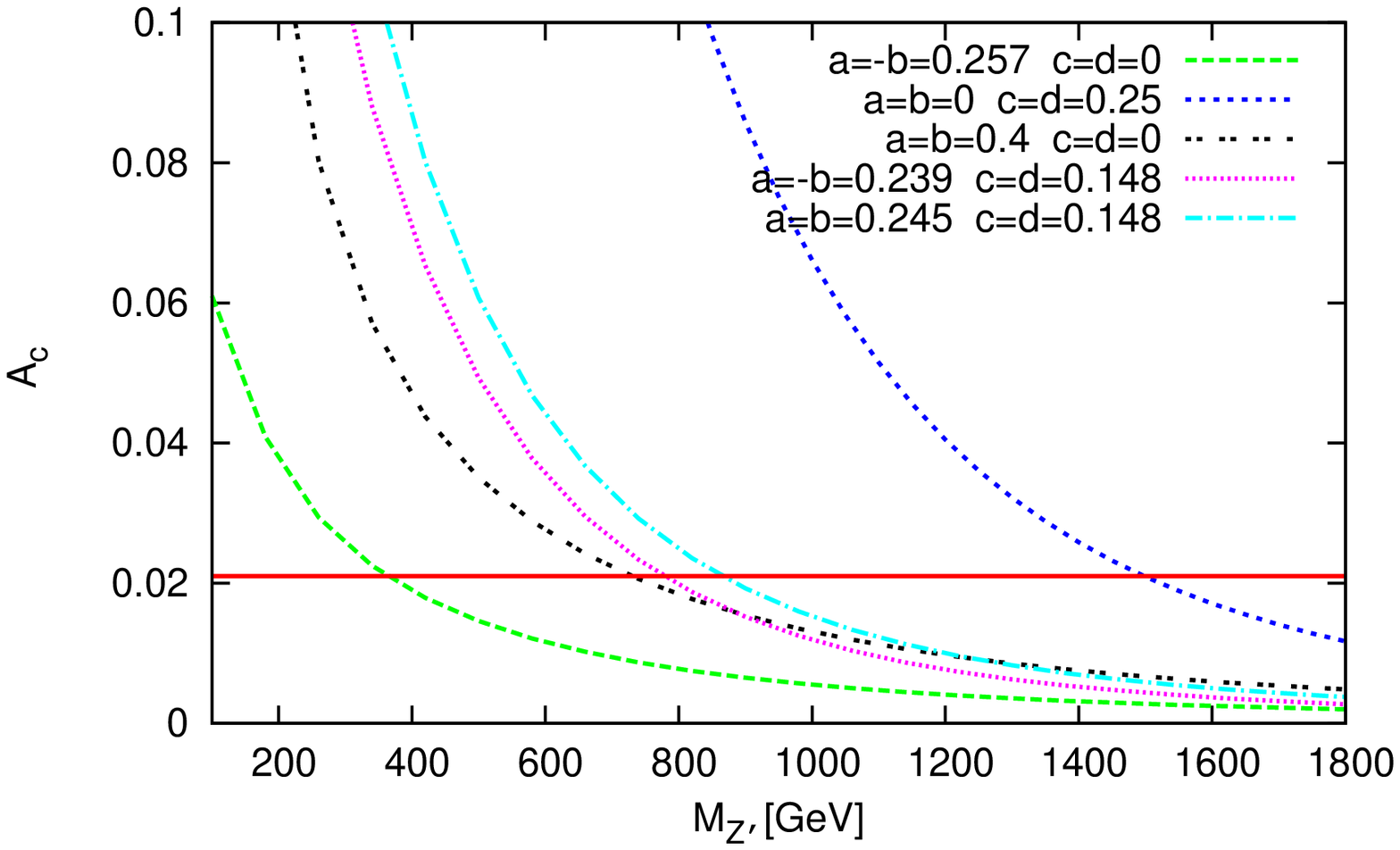,width=7.5cm}}
\centerline{\vspace{1.cm}\hspace{0.5cm}(a)\hspace{7cm}(b)}
\centerline{\vspace{-2.5cm}}
\end{center}
\caption{ The Top pair asymmetries as a function of  $Z'$ mass.
a) Forward-Backward asymmetry at Tevatron. b) Charge asymmetry at
LHC. The horizontal red lines show the allowed ranges of
experimental measurements for asymmetries.} \label{Asy}
\end{figure}

\begin{figure}
\begin{center}
\centerline{\hspace{3cm}\epsfig{figure=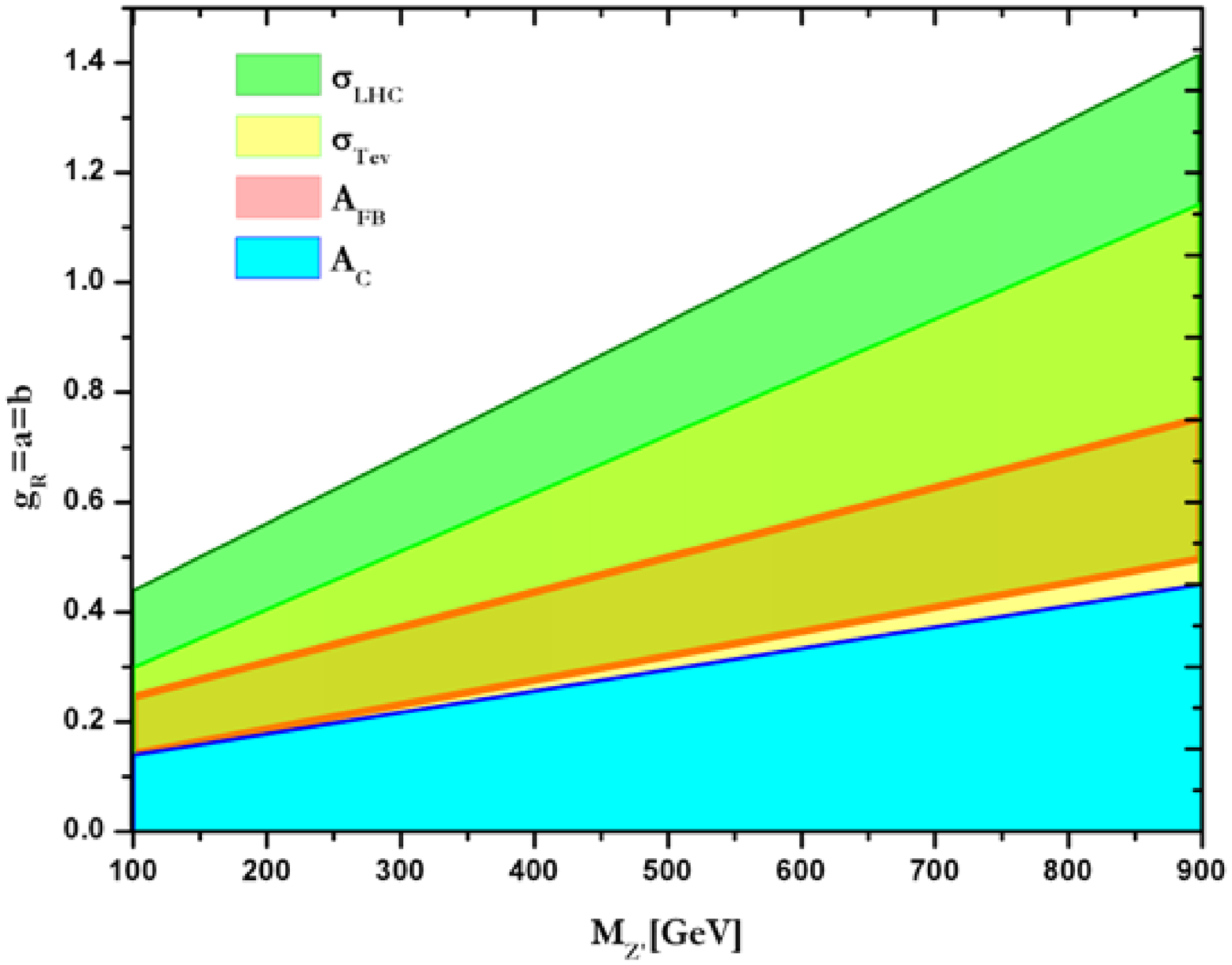,width=10.5cm}\hspace{-2.5cm}\epsfig{figure=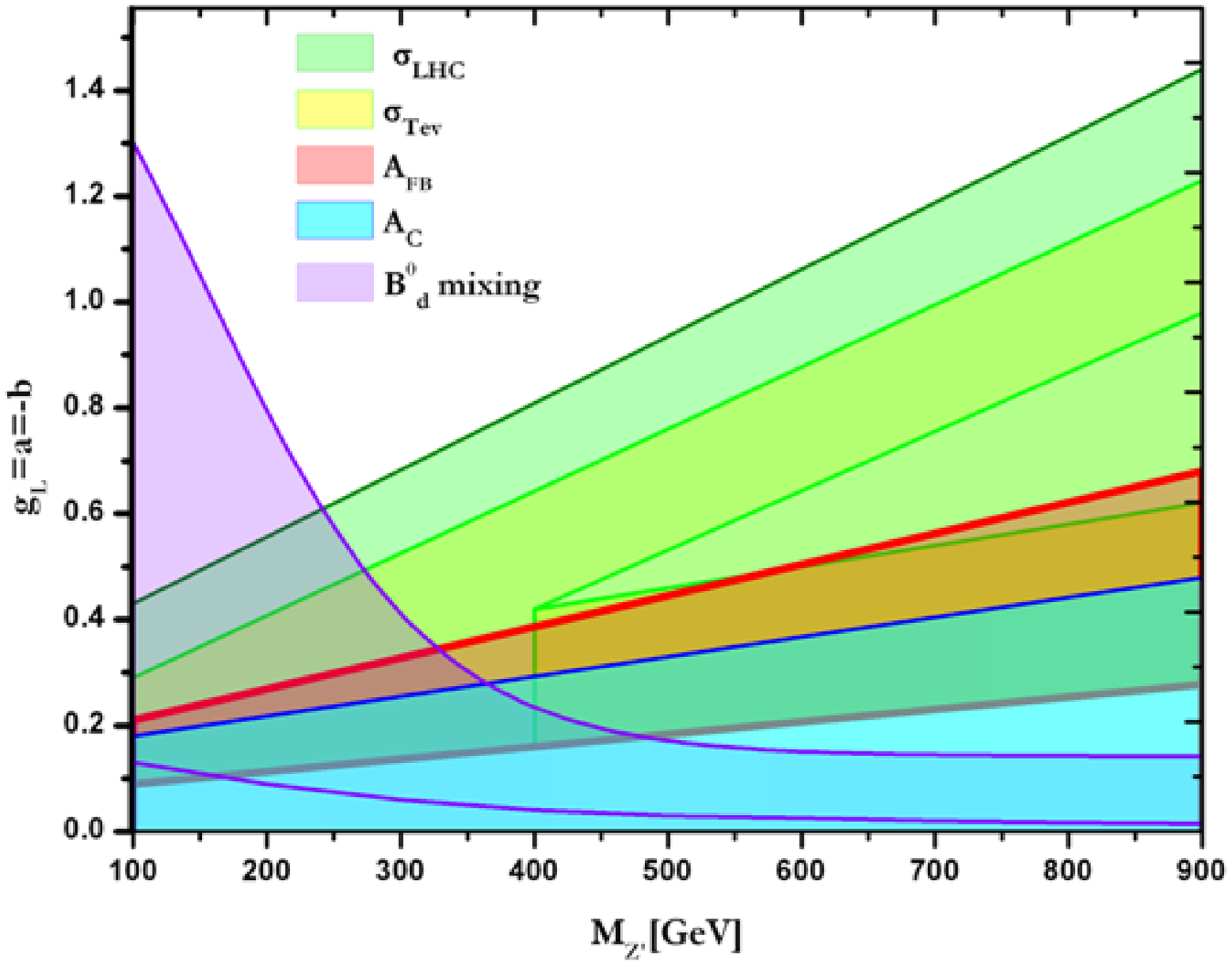,width=10.5cm}}
\centerline{\vspace{-2cm}}
\centerline{\hspace{3cm}\vspace{1mm}\hspace{-2.5cm}(a)\hspace{7cm}(b)}
\end{center}
\caption{Shaded areas depict ranges of parameters space in (a)
right-handed vector and (b) left-handed vector couplings and
$M_{Z'}$ plane for which are consistent with experimental
measurements and uncertainty of observables: $\sigma_{\rm LHC}$,
$\sigma_{\rm Tev}$, $A_{\rm FB}$, $A_{\rm C}$. Violet area in
Fig.~(b) shows allowed regions of $|g^L_{tu}|$ which are
consistent with experimental $B^0_d$ mixing constraints.}
\label{vector}
\end{figure}

\begin{figure}
\begin{center}
\centerline{\hspace{3cm}\epsfig{figure=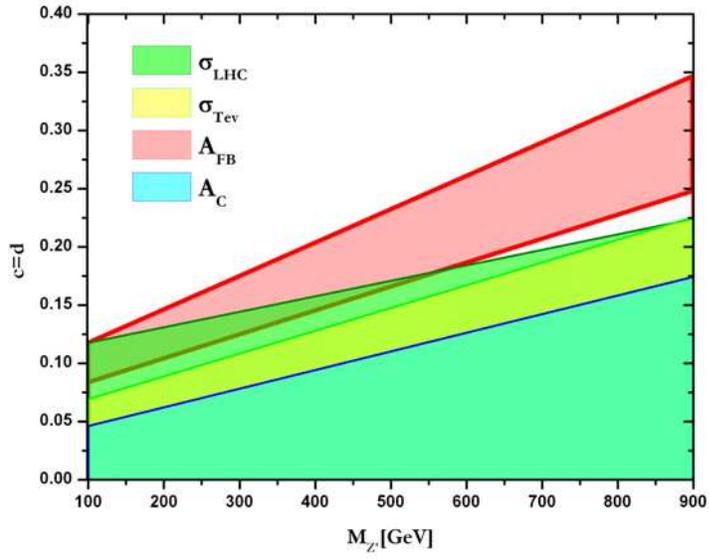,width=13.5cm}}
\centerline{\vspace{-2.5cm}}
\end{center}
\caption{Shaded areas depict ranges of parameters space in right-handed tensor
coupling and $M_{Z'}$ plane for which are consistent with
experimental measurements of observables: $\sigma_{\rm LHC}$,
$\sigma_{\rm Tev}$, $A_{\rm FB}$ and $A_{\rm C}$.} \label{tensor}
\end{figure}

\begin{figure}
\begin{center}
\centerline{\hspace{3cm}\epsfig{figure=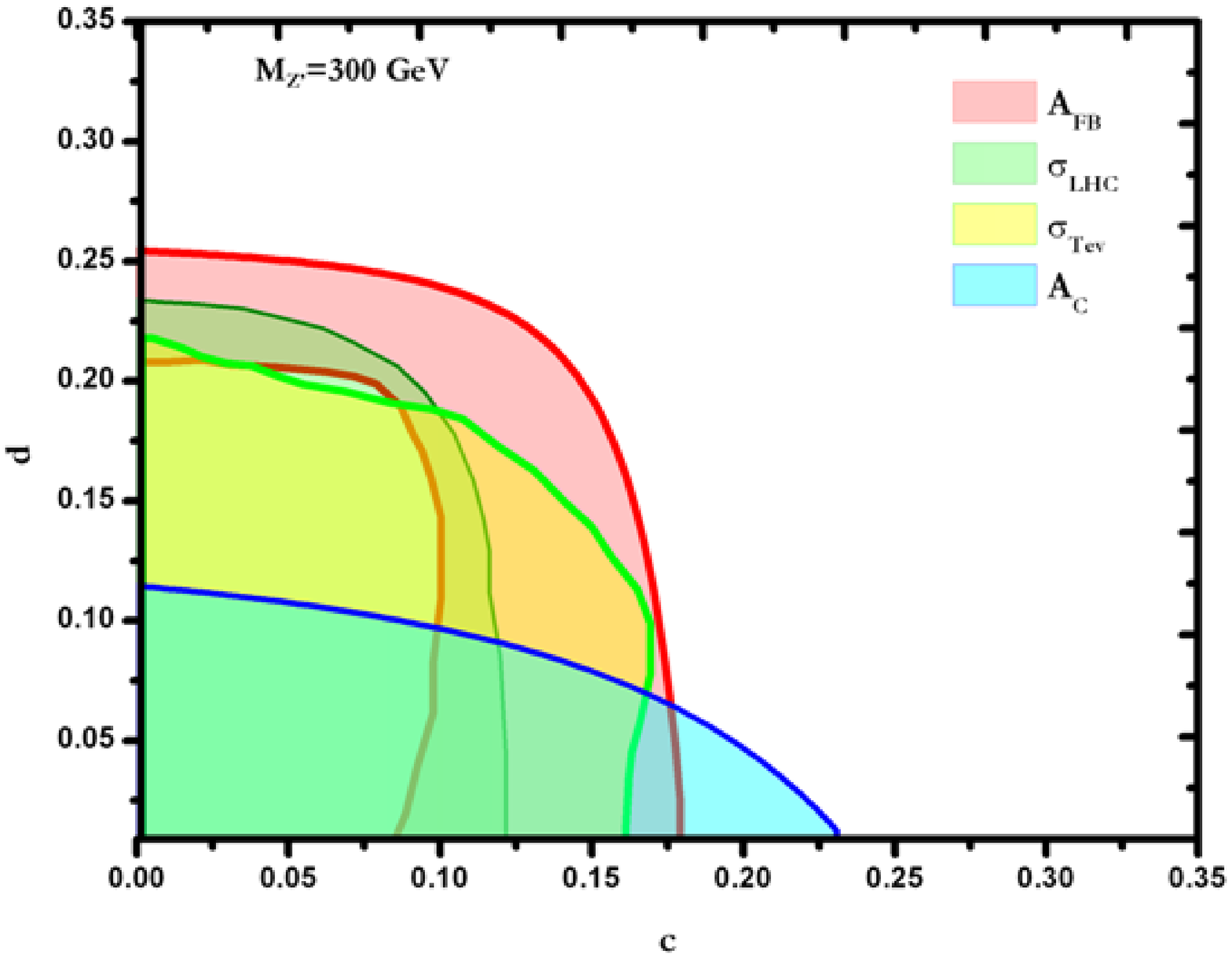,width=10.5cm}\hspace{-2.5cm}\epsfig{figure=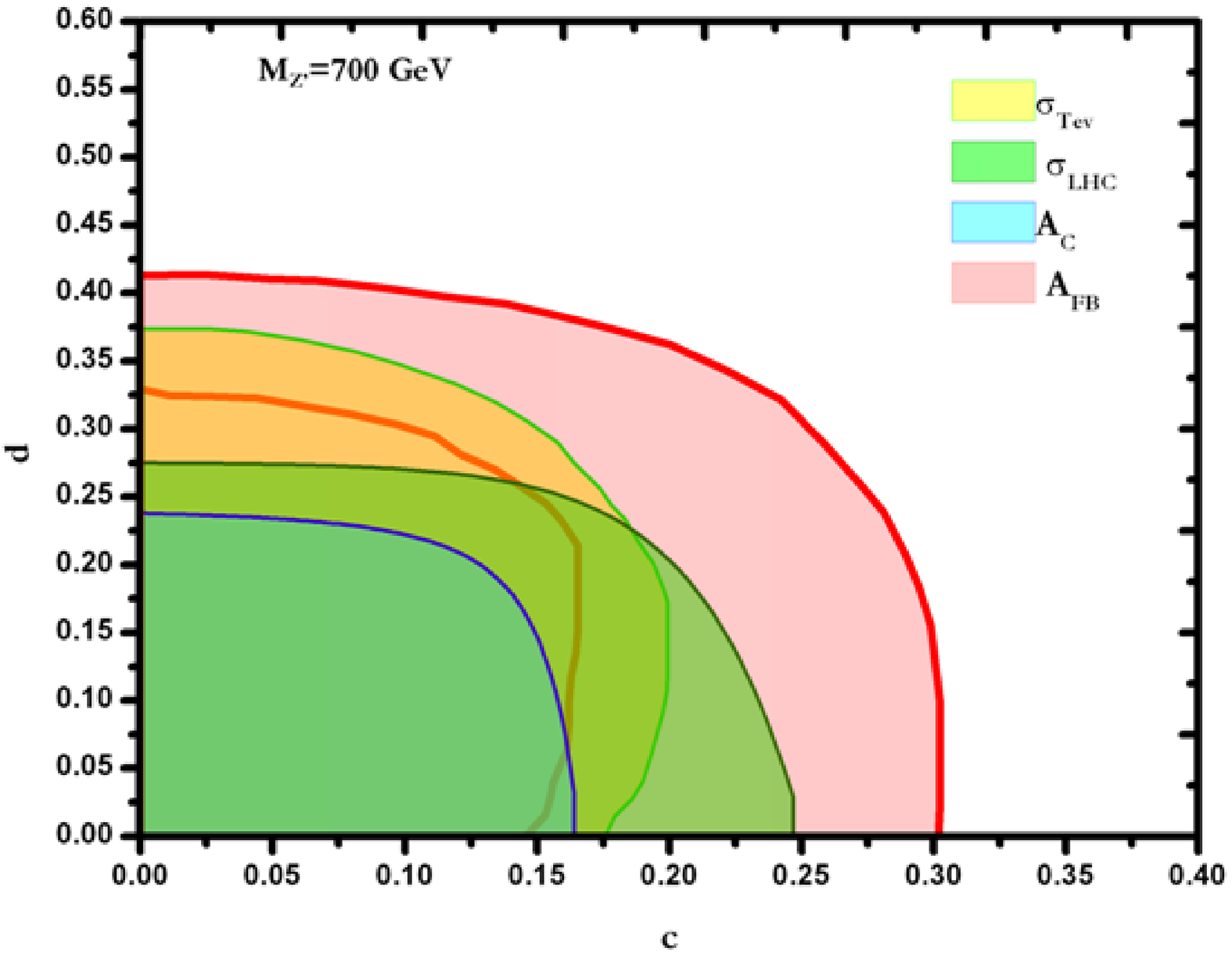,width=10.5cm}}
\centerline{\vspace{-2cm}}
\centerline{\hspace{3cm}\vspace{1.cm}\hspace{-2.5cm}(a)\hspace{7cm}(b)}
\centerline{\vspace{-2cm}}
\end{center}
\caption{Shaded areas depict ranges of parameters space in tensor
 couplings plane for which are consistent with
experimental measurements of observables: $\sigma_{\rm LHC}$,
$\sigma_{\rm Tev}$, $A_{\rm FB}$ and $A_{\rm C}$. In Fig.~(a) $M_{Z'}=300$ and in
Fig.~(b) $M_{Z'}=700$.} \label{tensor1}
\end{figure}

\begin{figure}
\begin{center}
\centerline{\hspace{3cm}\epsfig{figure=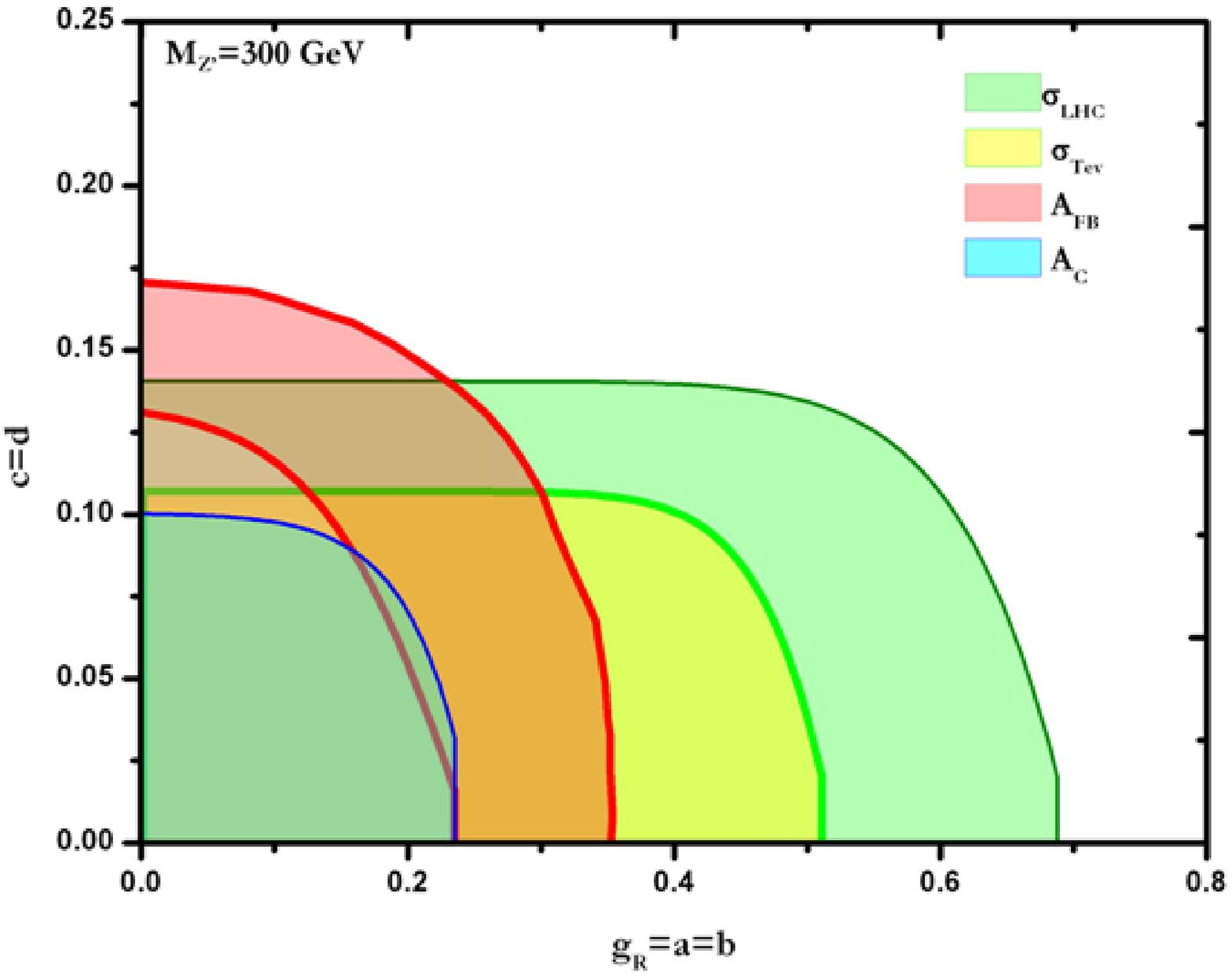,width=10.5cm}\hspace{-2.5cm}\epsfig{figure=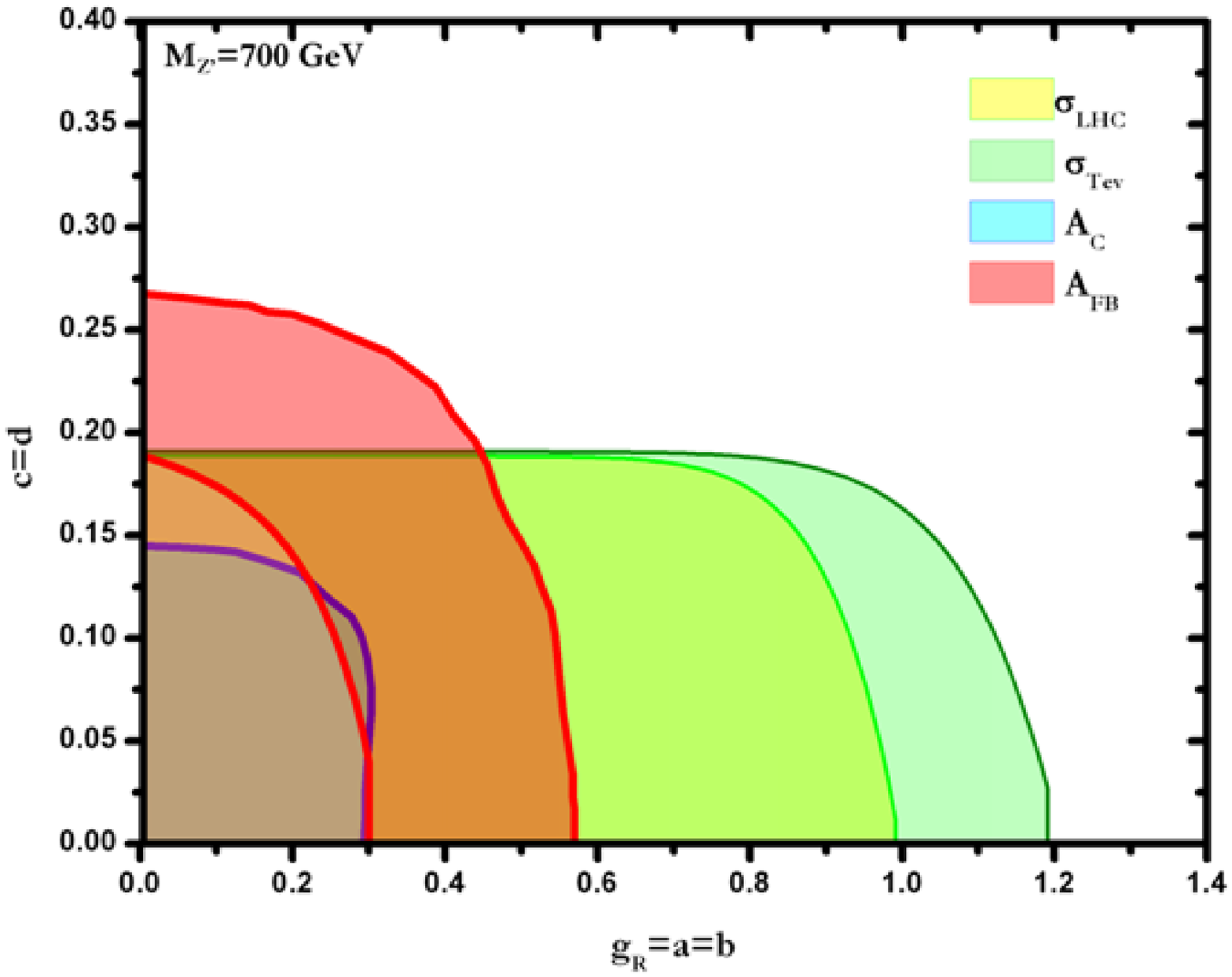,width=10.5cm}}
\centerline{\vspace{-2cm}}
\centerline{\hspace{3cm}\vspace{1.cm}\hspace{-2.5cm}(a)\hspace{7cm}(b)}
\centerline{\vspace{-2cm}}
\end{center}
\caption{Shaded areas depict ranges of parameters space in right-handed tensor
and  vector couplings plane for which are consistent with
experimental measurements of observables: $\sigma_{\rm LHC}$,
$\sigma_{\rm Tev}$, $A_{\rm FB}$ and $A_{\rm C}$. In Fig.~(a)
$M_{Z'}=300$ and in Fig.~(b) $M_{Z'}=700$} \label{gf1}
\end{figure}

\begin{figure}
\begin{center}
\centerline{\hspace{3cm}\epsfig{figure=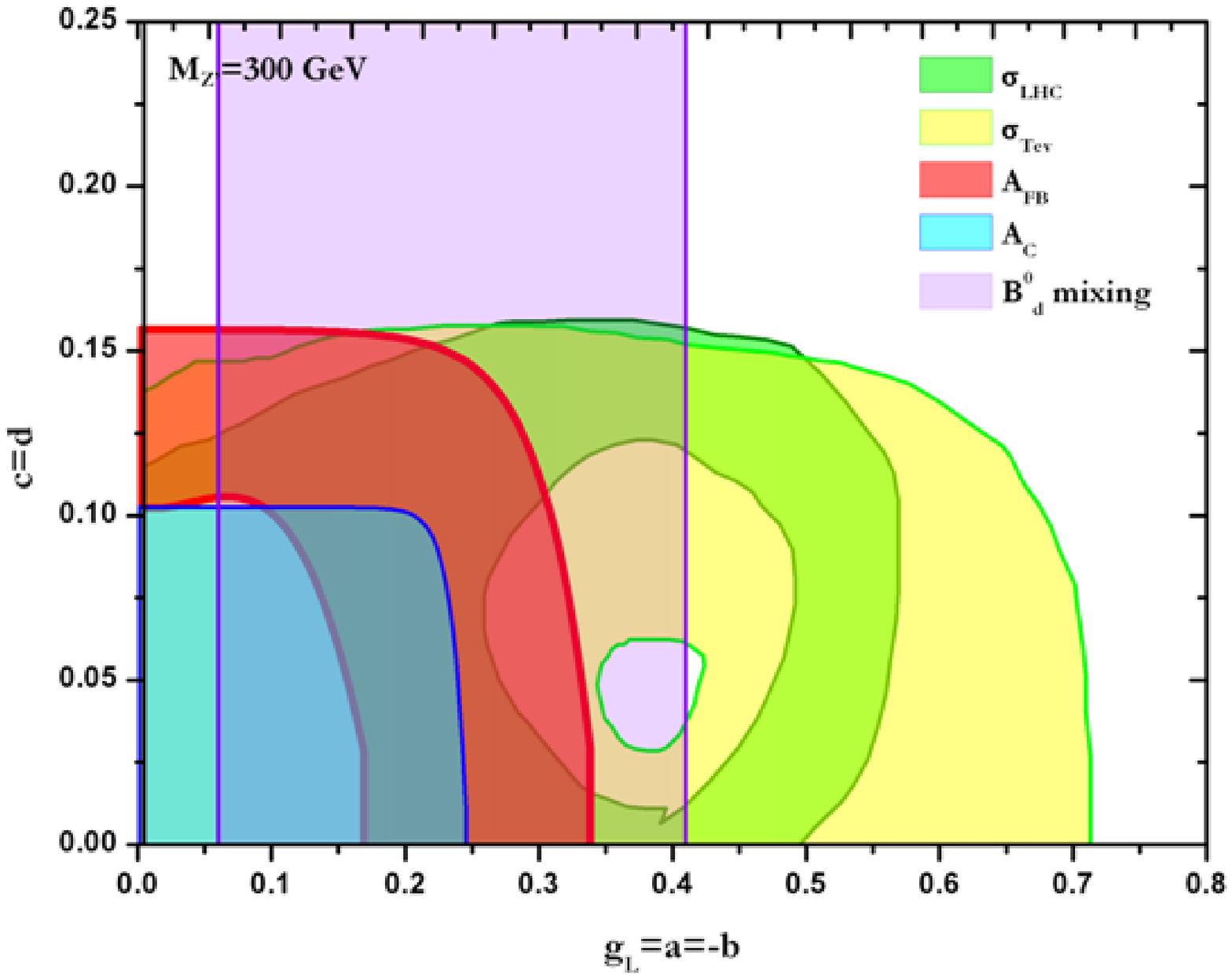,width=10.5cm}\hspace{-2.5cm}\epsfig{figure=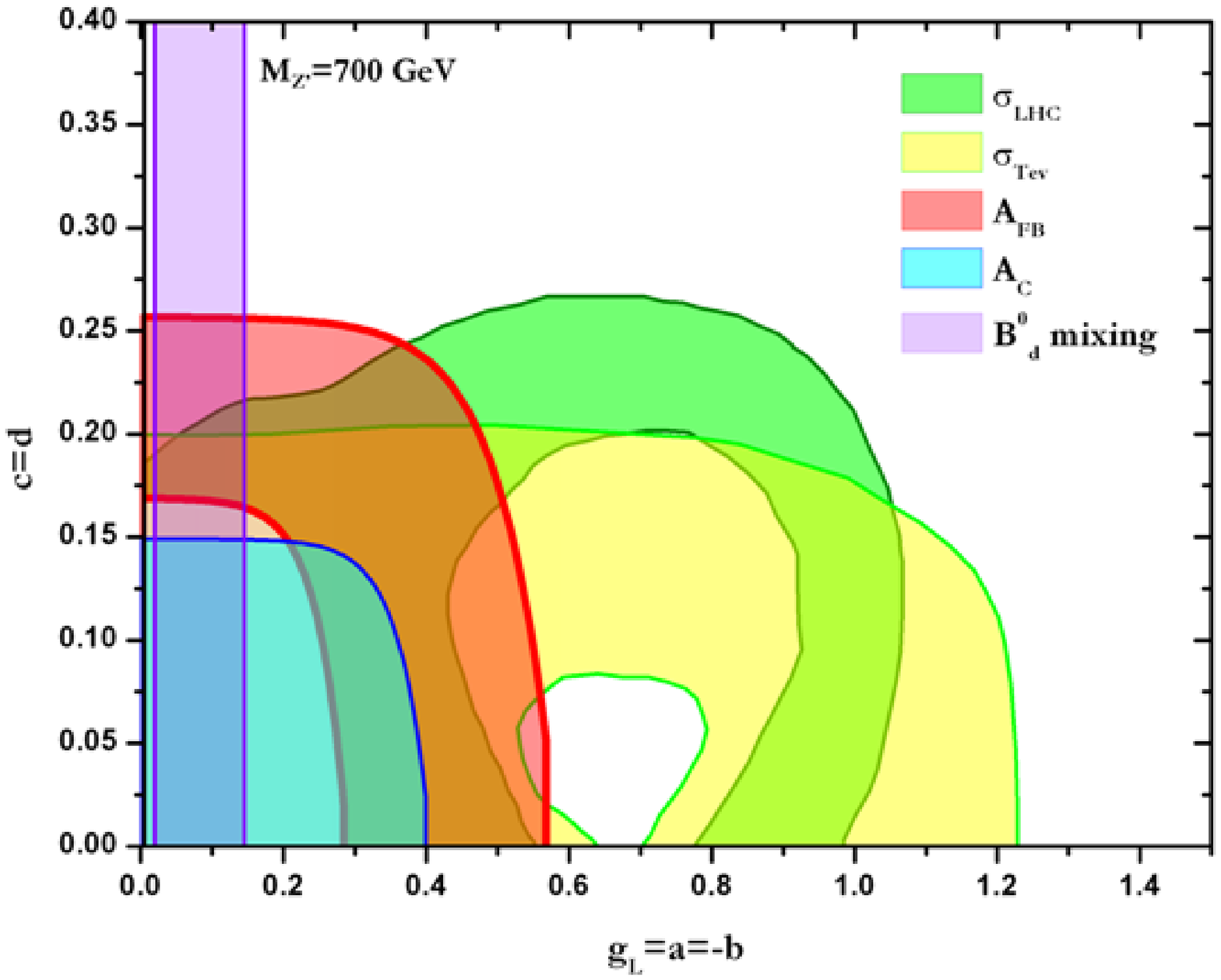,width=10.5cm}}
\centerline{\vspace{-2cm}}
\centerline{\hspace{3cm}\vspace{1.cm}\hspace{-2.5cm}(a)\hspace{7cm}(b)}
\centerline{\vspace{-2cm}}
\end{center}
\caption{Shaded areas depict ranges of parameters space in right-handed tensor
and left-handed vector couplings plane for which are consistent with
experimental measurements of observables: $\sigma_{\rm LHC}$,
$\sigma_{\rm Tev}$, $A_{\rm FB}$ and $A_{\rm C}$. Violet area
shows allowed regions of $|g^L_{tu}|$ which satisfy experimental
$B^0_d$ mixing constraints. In Fig.~(a) $M_{Z'}=300$ and in
Fig.~(b) $M_{Z'}=700$.} \label{gf2}
\end{figure}

Fig.~\ref{Asy}-a(b) depicts $A_{FB}$($A_c$) at Tevatron (LHC).
Different values for couplings have been considered according to
$B_q$ mixing constraints. As it can be seen, for instance the
allowed values for tensor couplings in $A_{FB}$($A_C$) curves are
satisfied for $600<M_{Z'}<800~\rm GeV$ ($M_{Z'}>1500~\rm GeV$).
This means, for given values $a=b=0$ and $c=d=0.25$, there is no
allowed region. To better study of all parameters space which
simultaneously satisfies experimental constraints on $\sigma_{\rm
Tevatron}$, $\sigma_{\rm LHC}$, $A_{FB}$ and $A_{C}$, we display
Figs.~\ref{vector}-\ref{gf2} and scan parameter space in these
categories:

\labelitemi\textbf{ $Z'$ with left-handed and right-handed vector
couplings (c=d=0):} In Fig.~\ref{vector}, shaded areas satisfy
experimental measurements of observables $\sigma_{\rm LHC}$,
$\sigma_{\rm Tev}$, $A_{\rm FB}$ and $A_{\rm C}$ in (a)
right-handed vector and (b) left-handed vector couplings and
$m_{Z}$ plane. As it was mentioned, there is no constraints on
right handed coupling which come from $B^0_q$ mixing. As it can
be seen, there is no overlapping area between $A_C$ and $A_{FB}$
allowable regions for right-handed coupling($a=b=g^R_{tu}$).
Therefore, measurement of $A_C$ at the LHC excludes any $Z'$ with
right-handed coupling which could explain $A_{FB}$ anomaly at
Tevatron.

In Fig.~\ref{vector}-b ($a=-b=g^L_{tu}$), we also consider $B^0_d$
mixing constraint on real part of left-handed coupling
$|g^L_{tu}|$ which have been taken from Fig.~2 of
\cite{Duraisamy:2011pt}. For this case, for real value of $g_L$,
we find regions that all experimental measurement are satisfied.
The constraints from $B^0_s$ mixing on $g^L_{tu}$ are suppressed
because the contribution of $g^L_{tu}$ to $B^0_s$ mixing is
proportional to $V^*_{us}V_{tb}$.

Nevertheless, $B^0_d$ mixing strongly constrains the left-handed
$tuZ'$ coupling. In this paper, we have assumed all couplings are
real. In \cite{Duraisamy:2011pt}, it is shown that for complex
coupling $g^L_{tu}$, $\rm Arg(g^L_{tu})$ must be between $-60~\rm
Deg$ and $-20~\rm Deg$. Therefore, there are no real $g^L_{tu}$
which satisfy the $B^0_d$ mixing constraints.

\labelitemi\textbf{ $Z'$ with pure tensor couplings (a=b=0):} In this case we can neglect the $B^0_q$ mixing constraints due to
suppressed effect of $m_b/m_t$ at the $b$ mass scale. In
Fig.~\ref{tensor}, we consider $Z'$ with right-handed tensor couplings ($c=d$).
 It is notable that allowed regions of top pair production cross section
at the LHC and Tevatron, overlap with allowed regions of $A_C$ and
$A_{FB}$. Nevertheless, there is no overlapping region between
the $A_C$ measurement at the Tevatron and $A_{FB}$ measurement at
the LHC. As a result, $A_C$ measurement can exclude $Z'$ gauge
boson with right-handed tensor coupling which could explain $A_{FB}$
anomaly measurement at Tevatron. For the case $c=-d$, the situation is similar to pervious case as expected because cross section is quadratically dependent on $d$ when $b=0$.

 For $Z'$ with general form of tensor coupling ($0\leq c\leq 1$ and $|d|\leq 1$), we display Fig.~\ref{tensor1}. It is
remarkable that there are overlapping regions between measured
$A_C$ at LHC and measured $A_{FB}$ at Tevatron. This means,
flavor changing $Z'$ gauge boson with general form of tensor coupling
could still explain forward-backward anomaly at Tevatron keeping all observables consistent with measurements. The comparison of Fig.~\ref{tensor1}-a and Fig.~\ref{tensor1}-b shows that the allowed region for $Z'$ will decline when $Z'$ mass increases.

In this figure, we consider situation that $0\leq c\leq 1$ and $0 \leq d \leq 1$. Similar condition exist for the case $-1 \leq d \leq 0$.

\labelitemi\textbf{  $Z'$ with general form of couplings:}
Fig.~\ref{gf1} depicts, allowed region of tensor axial and right-handed
couplings for which experimental measurements of observables
$\sigma_{\rm LHC}$, $\sigma_{\rm Tev}$, $A_{\rm FB}$ and $A_{\rm
C}$ are satisfied. As it was mentioned, for right-handed
coupling, the contributions of $tuZ'$ vertex to $B^0_d$ and
$B^0_s$ mixing are suppressed by a factor of $m^2_u/m^2_W$. As it has been shown in these figures,
there are small overlapping regions between the measured
$A_C$ at the LHC and the measured $A_{FB}$ at the Tevatron. For the case $a=b=1$ and $(c\neq |d|)\leq 1$, we expect that similar to pure tensor coupling,
there are allowed regions of $A_C$ and $A_{FB}$. In
Fig.~\ref{gf2}, we consider $Z'$ gauge boson with left-handed
vector and tensor couplings. For this case, we have taken
constraints $B_d$ mixing from \cite{Duraisamy:2011pt}. Violet
area in Fig.~\ref{gf2} shows allowed regions for $|g^L_{tu}|$
which satisfy experimental $B^0_d$ mixing constraints. 
In this paper, we suppose  $g^L_{tu}$ is real and Fig.~\ref{gf2}-a shows that for real value of $g^L_{tu}$, there is a overlapping region which satisfy all experimental constraints. As it was mentioned (in Fig~.\ref{vector}-b) for the case $a=-b=g^L_{tu}$, $B_d$ mixing can strongly constrain the $tuZ'$ coupling. In fact there are no real $g^L_{tu}$ consistent with $B^0_d$
mixing constraints. This means that if we relax $B^0_d$ constraints
on phase of $g^L_{tu}$ \cite{Duraisamy:2011pt}, there are allowed
regions for $A_C$ and $A_{FB}$ measurements. Nevertheless these
values for couplings can not satisfy $B_d$ mixing constraints. The case which $a=b=1$ and $(c\neq |d|)\leq 1$ is excluded by experimental $B^0_d$ mixing constraints.

\section{Concluding remarks}
In this paper, we have studied the effects of $Z'$ gauge boson on
top pair asymmetries and its cross section productions at the
Tevatron and the LHC. It was recently  shown that measurement of
charge asymmetry of top pair events at the LHC excludes any flavor
violating $Z'$ vector gauge boson in all of parameters
space\cite{Fajfer:2012si}. We have focused on the flavor changing
$Z'$ gauge boson model with general form of the couplings
including left-handed, right-handed vector and tensor couplings.
We have also discussed the effect of $tuZ'$ couplings on $B^0_q$
($q=d,s$) mixing and considered consistent couplings with $B^0_d$
constraints. We have shown that right-handed vector $tuZ'$
coupling, there is no overlapping region which satisfies
simultaneously  $A_C$ and $A_F$ measurements. For left-handed
couplings, these regions exist but $B^0_d$ mixing constraint do
not allow us to consider these couplings. Similar conditions exist
for $tuZ'$ left-handed and tensor couplings with specific chirality.

The main point is that if we consider general form of tensor coupling, all experimental measurements including top pair
asymmetries and top pair total cross section at the Tevatron and the LHC
are satisfied and also $B^0_q$ mixing does not limit the
couplings. For right-handed vector and
tensor couplings,  there is small region in parameters space which satisfied all experimental bounds. This means flavor changing $Z'$ gauge boson still can
explain forward-backward anomaly at Tevatron.

\section{Acknowledgement}
The author would like to thank   M. Mohammadi Najafabadi  for the
useful discussions. He is also grateful to S. Paktinat for
careful reading of the manuscript and the useful remarks.

\end{document}